\documentclass[twocolumn,a4paper,showpacs,preprintnumbers,amsmath,amssymb,nofootinbib,floatfix]{revtex4}
\input psfig.sty  
\def \be {\begin{equation}} 

\def \ee {\end{equation}} 
\def \bea {\begin{eqnarray}} 
\def \eea {\end{eqnarray}} 

\usepackage{graphicx}
\usepackage{dcolumn}
\usepackage{bm}
\usepackage{epsfig} 

\begin{document}
\title{Probing the cosmic distance duality with strong gravitational lensing\\ and supernovae Ia data}
\author{R. F. L. Holanda$^{1}$  } \email{holanda@uepb.edu.br}
\author{V. C. Busti$^{2}$} \email{vcbusti@astro.iag.usp.br}
\author{J. S. Alcaniz$^3$}\email{alcaniz@on.br}
\affiliation{$^1$ Departamento de F\'{\i}sica, Universidade Estadual da Para\'{\i}ba, 58429-500, Campina Grande - PB, Brasil,
\\ and Departamento de F\'{\i}sica, Universidade Federal de Campina Grande, 58429-900, Campina Grande - PB, Brasil
\\$^2$ Departamento de F\'{\i}sica Matem\'{a}tica, Instituto de F\'{\i}sica, Universidade de S\~{a}o Paulo, 
CP 66318, CEP 05508-090, S\~{a}o Paulo - SP, Brasil
\\ $^3$ Departamento de Astronomia, Observat\'orio Nacional, 20921-400, Rio de Janeiro - RJ, Brasil}

\date{\today}

\begin{abstract}

We propose and perform a new test of the cosmic distance-duality relation (CDDR), $D_L(z) / D_A(z) (1 + z)^{2} =  1$, where $D_A$ is the angular diameter distance and $D_L$ is the luminosity distance to a given source at redshift $z$, using strong gravitational lensing (SGL) and  type Ia Supernovae (SNe Ia) data. We show that the ratio $D=D_{A_{12}}/D_{A_2}$ and $D^{*}=D_{L_{12}}/D_{L_{2}}$, where the subscripts 1 and 2 
correspond, respectively, to redshifts $z_1$ and $z_2$, are linked by $D/D^*=(1+z_1)^2$ if the CDDR is valid. We allow departures 
from the CDDR by defining two functions for $\eta(z_1)$, which equals unity when the CDDR is valid.  
We find that combination of SGL and SNe Ia data favours no 
violation of the CDDR at 1$\sigma$ confidence level ($\eta(z) \simeq 1$), in complete agreement with other tests and reinforcing
the theoretical pillars of the CDDR.
\end{abstract}
\pacs{98.80.-k, 95.36.+x, 98.80.Es}
\maketitle

\section{Introduction}

For the first time in the history of cosmology, the variety and quality of current cosmological data provide the possibility of testing some fundamental hypotheses of the standard cosmological model. One of these hypotheses is the validity of the so-called cosmic distance-duality relation (CDDR), which is derived from Etherington reciprocity theorem (Etherington, 1933). The CDDR, expressed as 
\begin{equation}
{D_L(z) \over  D_{A}(z) (1 + z)^{2}} =  1\;,
\end{equation}
where $D_A(z)$ is the angular diameter distance and $D_L(z)$ is the luminosity distance to a given source at redshift $z$, holds if photons follow null (unique) geodesic and the geodesic deviation equation is valid, along with the assumption that the number of photons is conserved over the cosmic evolution (Bassett \& Kunz, 2004). The CDDR plays an essential role in observational cosmology and any departure from it could point to exciting new fundamental physics or unaccounted systematic errors (for a general discussion, see Ellis, 1971).

In recent years, several authors have proposed methods to test the CDDR. Generally speaking, we can roughly divide them in two classes, cosmological model-dependent tests, usually performed within the cosmic concordance $\Lambda$CDM model, and model-independent ones. For instance, Uzan et al. (2005) showed that the combination of the Sunyaev-Zeldovich effect (SZE) plus X-ray observations, used to measure angular diameter distance  to galaxy clusters, is dependent on the CDDR validity. By 
using 18 measurements of $D_A(z)$ to galaxy clusters described by a spherical model and the $\Lambda$CDM concordance model these 
authors showed that the CDDR validity was only marginally verified. Holanda {\it{et al.}} (2010) and 
Meng {\it{et al.}} (2012), also using SZE/X-ray observations of galaxy clusters,  showed that triaxial ellipsoidal 
model is a better geometrical hypothesis describing the structure of the galaxy cluster compared with the 
spherical model if the CDDR is valid in cosmological observations.  Avgoustidis {\it{et al.}} (2010; 2012) adopted a modified expression of the CDDR, 
$D_L(z)/ D_A(z)(1+z)^{2+\epsilon} = 1$,  to constrain, in the context of a flat $\Lambda$CDM model, the cosmic 
opacity  from type Ia supernova  (SNe Ia) and $H(z)$ data. They found $ \epsilon= -0.04^{+0.08}_{-0.07}$ ($2\sigma$). 
In this line, Holanda \& Busti (2014) probed the  cosmic opacity at high redshifts with gamma-ray bursts and $H(z)$ 
data and showed that this combination of data is compatible with a transparent universe ($\epsilon = 0$) at $1\sigma$ level, regardless of the dark energy equation-of-state parameter $w$ assumed in the analysis. 

On the other hand, several cosmological model-independent tests for the CDDR have also been proposed. 
De Bernardis, Giusarma \& Melchiorri (2006), Holanda {\it{et al.}} (2011) and Li, Wu \& Yu 2011 confronted $D_A(z)$ measurements to galaxy clusters with SNe Ia data obtaining that the CDDR is verified at 1$\sigma$ when the galaxy clusters are 
described by ellipsoidal model. By showing that 
the X-ray gas mass fraction ($f_{X-ray}$) of galaxy clusters also depends on the CDDR validity, Gon\c{c}alves {\it{et al.}} (2012) 
proposed a test involving  samples of gas mass fractions and SNe Ia observations.
Another test using exclusively gas mass fractions was proposed by Holanda {\it{et al.}} (2012). These 
authors showed  that the relation between $f_{X_ray}$ and $f_{SZE}$ observations is 
given by $f_{SZE}=\eta f_{X−ray}$, where $\eta$ quantifies departures of the CDDR validity. No violation was found. It was shown no violation 
of the CDDR. Santos-da-Costa {\it{et al.}} (2015), applying gaussian process,  proposed a test based on galaxy clusters observations and $H(z)$ measurements and no evidence of deviation of the CDDR validity was found. By using measurements of the cosmic microwave background black-body spectrum, Ellis {\it{et al.}} (2013) showed 
that the reciprocity relation 
cannot be violated by more than $0.01\%$ between decoupling and today. 
Recently, Liao {\it{et al.}} (2015) introduced a new method to test the CDDR based on strong gravitational lensing systems (Cao et al. 2015) and the most recent SNe Ia compilation (Betoule et al. 2014). 

Although most of the results so far have been consistent with the validity of the CDDR, new methods with different astronomical 
observations and redshift range offer a path to validate the whole
cosmological framework as well as to detect unexpected behaviour or systematic errors.  
Therefore, in this paper,  we propose  a  new  test  which uses the strong 
gravitational lensing effect and  luminosity distances from type Ia supernovae. The paper is organised as follows.  In Section 2 we describe our new method to test the CDDR, while in Section 3 the observational quantities used in this work are discussed.  The corresponding constraints on the departures of the  CDDR  are investigated and discussed in Section 4.  We end this paper by summarizing our main results in Section 5.

\section{Method}

\subsection{Gravitational Lensing}

Strong gravitational lensing (SGL) is a powerful astrophysical effect to probing both gravity theories and cosmology. It occurs whenever the 
source ($s$), the lens ($l$) and the observer ($o$) are so well aligned that the observer-source direction lies 
inside the so-called Einstein radius of the lens (for a complete discussion on this effect, see Schneider, Ehlers \& Falco, 1992). In a cosmological context the source is usually a galaxy or quasar with a 
galaxy acting as the lens. Recent papers have explored strong gravitational lensing  systems for measuring cosmological parameters (Biesiada, Malec, \& Piorkowska 2011; Yuan \& Wang 2015, Cao et al. 2015).

The Einstein radius $\theta_E$, under assumption of the singular isothermal sphere (SIS) model for the lens, is given by
\begin{equation}
 \label{thetaE_SIS}
{{\theta}_E}=4{\pi}{\frac{D_{A_{ls}}} {D_{A_s}}}{\frac{{\sigma}^2_{SIS}} {c^2}},
\end{equation}
where $c$ is the speed of light, $D_{A_{ls}}$ and $D_{A_{s}}$ are the angular diameter distances between lens-source and source-observer, respectively, and $\sigma_{SIS}$ is the velocity dispersion due to lens mass distribution.{$\sigma_{SIS}$  is not to be exactly equal to the observed stellar velocity dispersion $\sigma_0$, there is a strong indication that dark matter halos are dynamically hotter than
the luminous stars based on X-ray observations.} In order to account for this difference  it is introduced a phenomenological free parameter
$f_e$, defined by the relation: $\sigma_{SIS} = f_e\sigma_0$, where $(0.8)^{1/2} < f_e < (1.2)^{1/2}$ (Ofek et al. 2003). 

In our CDDR test we are interested in the ratio between angular distances: 
\begin{equation}
D=D_{A_{ls}}/D_{A_{s}}=\frac{\theta_E c^2}{4 \pi \sigma^2_{SIS}}.
\end{equation}
If one assumes the CDDR validity it is possible to transform the above expression into a luminosity 
distance ratio. For the angular diameter distance of the source $s$ we have
\begin{equation}
 D_{L_{s}}=(1+z_s)^2 D_{A_{s}}\;,
\end{equation}
whereas the luminosity and angular distances between lens and source are linked by (Bartelmann \& Schneider, 2001)
\begin{equation}
 D_{L_{ls}}=\frac{(1+z_s)^2}{(1+z_l)^2}D_{A_{ls}}.
\end{equation}
From the above expressions, one may show that
\begin{equation}
D/D^*=(1+z_l)^2\;,
\label{razao}
\end{equation}   
where $D*=D_{L_{ls}}/D_{L_{s}}.
$
Therefore, obtaining $D^*$ from observations, the expression (\ref{razao}) can be modified 
to investigate  the CDDR validity. In order to obtain $D^*$ it is necessary to know $D_{L_{s}}$ at the same redshift of source in the lensing system and $D_{L_{ls}}$, the luminosity distance between lens and source.

\begin{figure*}
\centering
\includegraphics[width=0.47\textwidth]{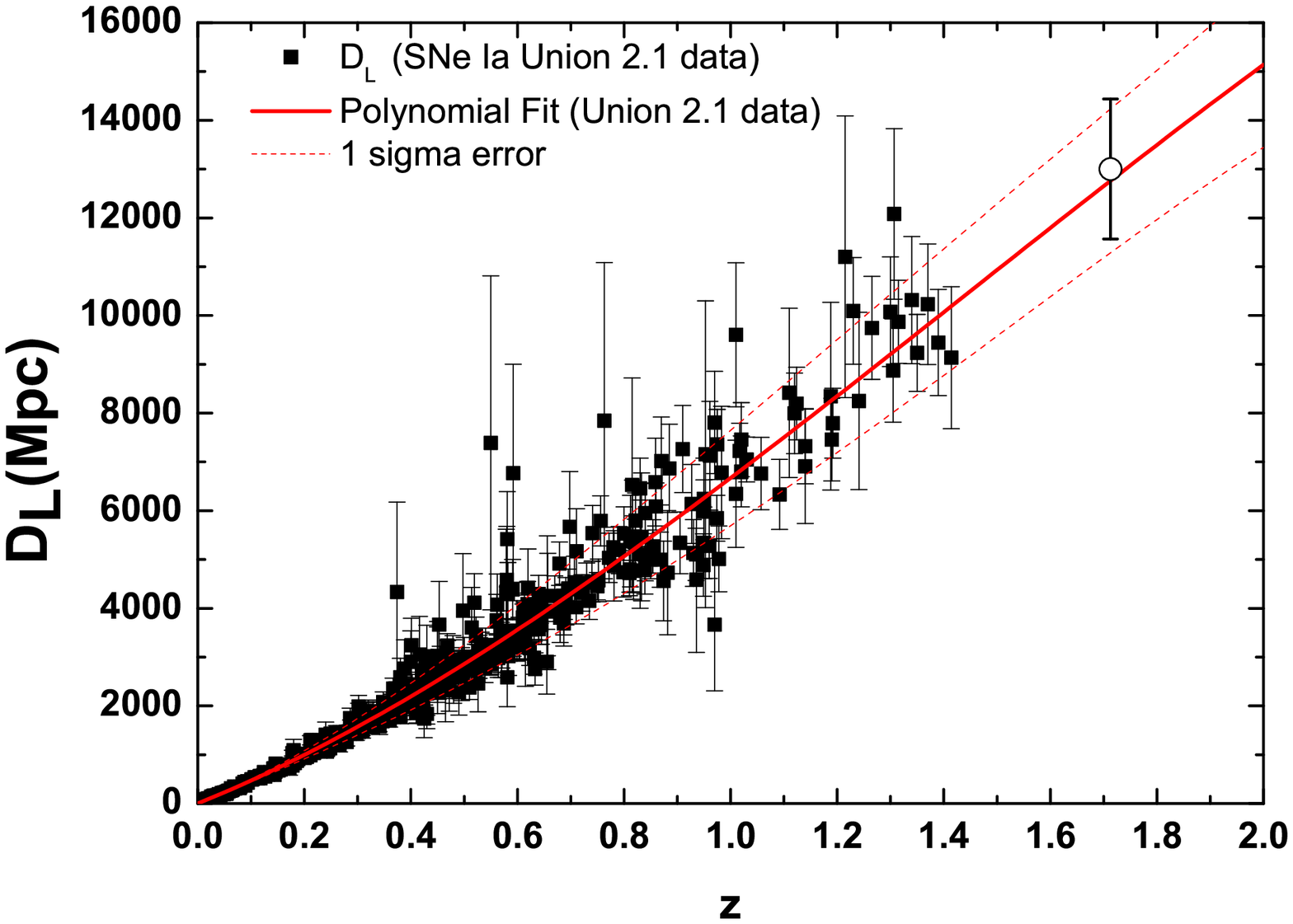}
\hspace{0.3cm}
\includegraphics[width=0.47\textwidth]{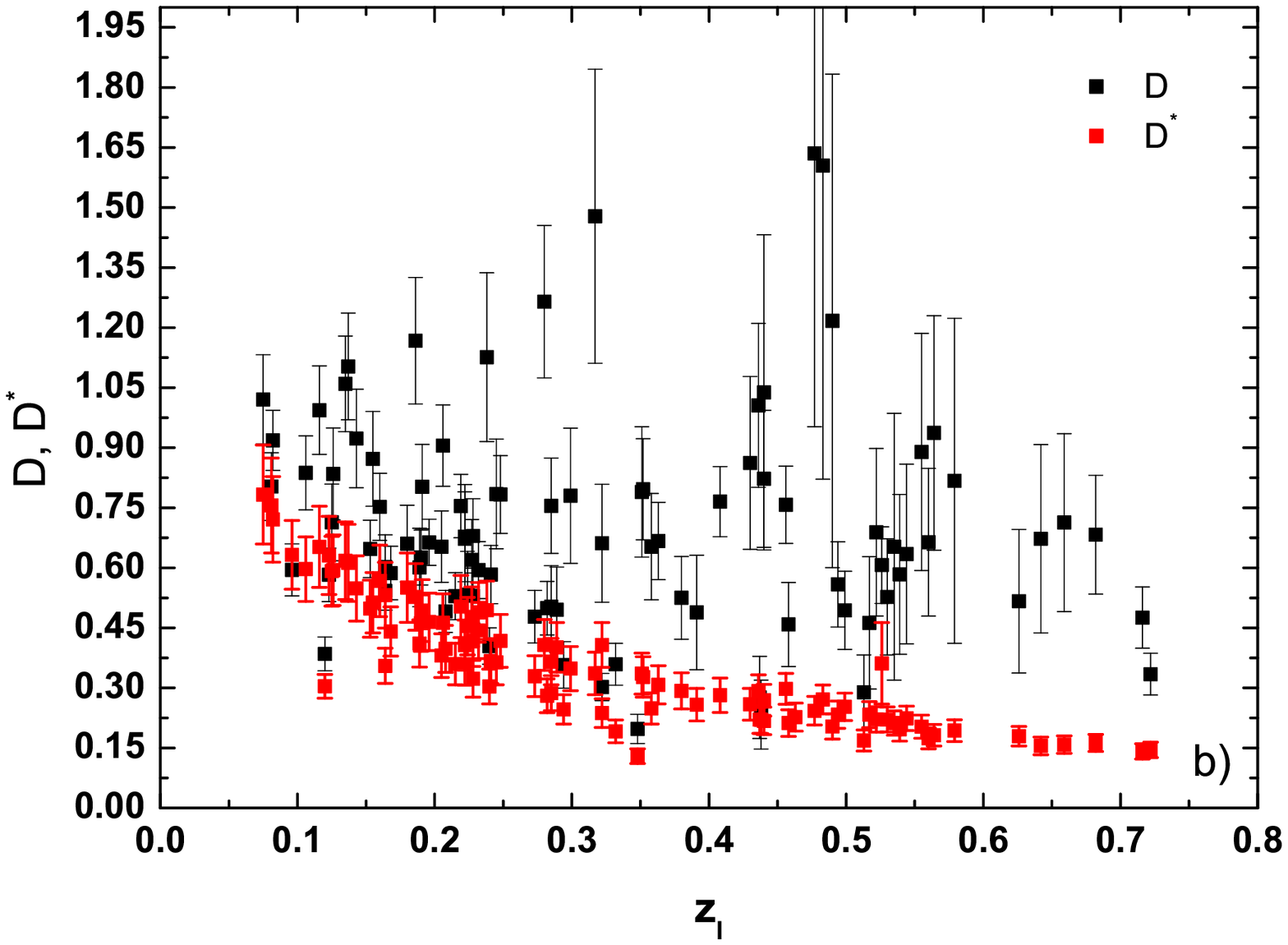}
\caption{(a) Measurements of $D_{\scriptstyle L}$ extracted from the Union2.1 SNe Ia sample (filled black squares). 
The curves stand for the  polynomial fit of $D_{\scriptstyle L}$ points from SNe Ia 
data and the corresponding 1$\sigma$ error. The open circle corresponds to the most distant 
spectroscopically confirmed SNe Ia ($z=1.713$). (b) The quantities $D$ and $D^*$ obtained from gravitational lensing (SIS model with $f_e=1$) and SNe Ia data, assuming the Planck best-fit $\Lambda$CDM model.
}
\end{figure*}

\subsection{Luminosity distance}

In our work $D_{L_s}$ is obtained as follows: we first estimate the luminosity distance $D_{L}$  for each SNe Ia of the sample by using 
their distance modulus measurements. The relation between the distance modulus $\mu$ and the luminosity distance is:
\begin{equation}
\mu_B(z) = m_B - M_B = 5 \log_{10} \left(\frac{D_L(z)}{1Mpc}\right)
+25,
\end{equation}
where $M_B$ is the absolute magnitude of the source and $m_B$ is the apparent magnitude ($B$ is for the $B$-band). { It is important to 
stress that the distance moduli were obtained fitting some SNe Ia light-curve parameters in a flat $\Lambda$CDM model, which makes them model
dependent. However, this dependence is much smaller than the errors coming from the gravitational lensing modelling, so it does not 
affect our results strongly.} Applying the transformation from distance modulus to luminosity distance,  we perform a  
polynomial fit to $D_{L}$ of the SNe I data (see Fig. (1a)  and the discussion in the next section). Thus, we can calculate $D_{L_s}$ at 
each point of interest (source in the lensing system). Previous papers used pairs of  SNe Ia with some other astronomical object whose 
the redshifts difference between them were smaller than $\Delta z \simeq 0.005$. The advantage of using polynomial fit on the SNe Ia data 
is to avoid any redshift difference as well as to minimize the bias from the objects in the same redshift but in different direction on the sky.

The distance $D_{L_{ls}}$, the luminosity distance  between lens and source, is obtained by using two cosmological models: flat  $\Lambda$CDM concordance model from the Planck collaboration  (Ade et al. 2015) and flat $\omega(z)$CDM 
from WMAP9 satellite results (Hinshaw et al. 2013). The $D_{L_{ls}}$ expressions for these models are 
\begin{equation}
 D_{L_{ls}}=\frac{(1+z_s)}{(1+z_l)^2}\frac{c}{H_0}\int_{z_l}^{z_s}\frac{dz}{E(z)},
\label{distance}
\end{equation}
where $E(z)$ for $\Lambda$CDM model is
\begin{equation}
\sqrt{\Omega_M(1+z)^3 + 1 - \Omega_{M}},
\end{equation}
and for $\omega(z)$CDM is
\begin{equation}
\sqrt{\Omega_M(1+z)^3 + (1 - \Omega_{M})(1+z)^{3(1+\omega_0+\omega_a)}e^{3\omega_a(\frac{1}{1+z}-1)}}\;,
\end{equation}
where the dark energy equation-of-state parameter is assumed to be given by $\omega(z)= \omega_0 + \omega_az/(1+z)$. 

Combining temperature and lensing data, the Planck collaboration found, for a flat $\Lambda$CDM universe, $H_0=67.8 \pm 0.9$ and $\Omega_M=0.308 \pm 0.012$ $(1\sigma)$. For a flat  $\omega(z)$CDM model, we use the following values to parameters at  1$\sigma$ obtained from WMAP 9 years (Hinshaw et al. 2013): $H_0=71 \pm 1.3$, $\omega_0=-1.17 \pm 0.13$, $\omega_a= 0.35 \pm 0.50$  and $\Omega_M=0.274 \pm 0.011$. In Fig. (1b) we plot $D$ and $D^*$ using the gravitational lensing and SNe Ia data. For simplicity, we only show these quantities for the $\Lambda$CDM scenario.

\begin{figure*}
\centering
\includegraphics[width=0.48\textwidth]{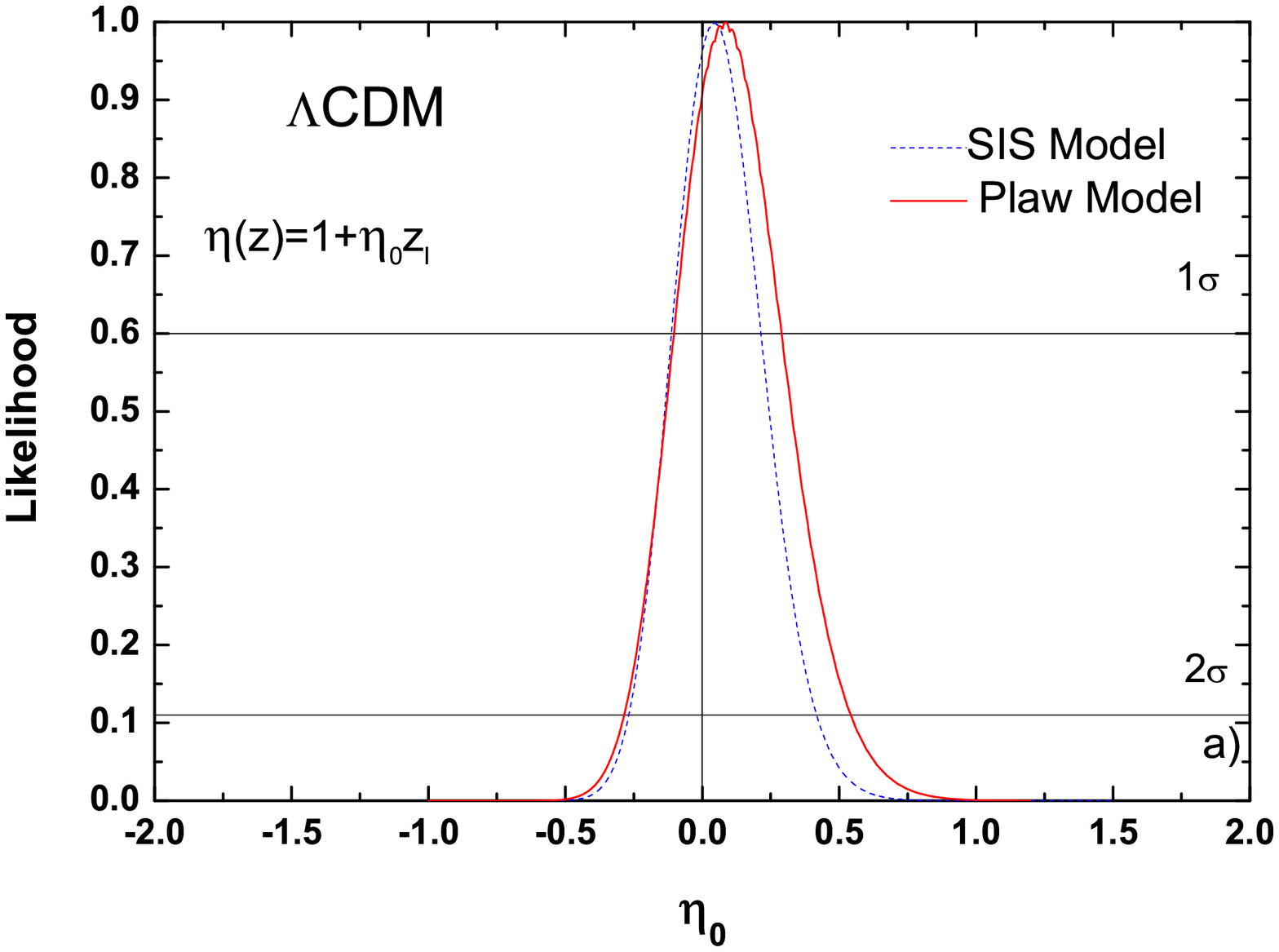}
\hspace{0.3cm}
\includegraphics[width=0.48\textwidth]{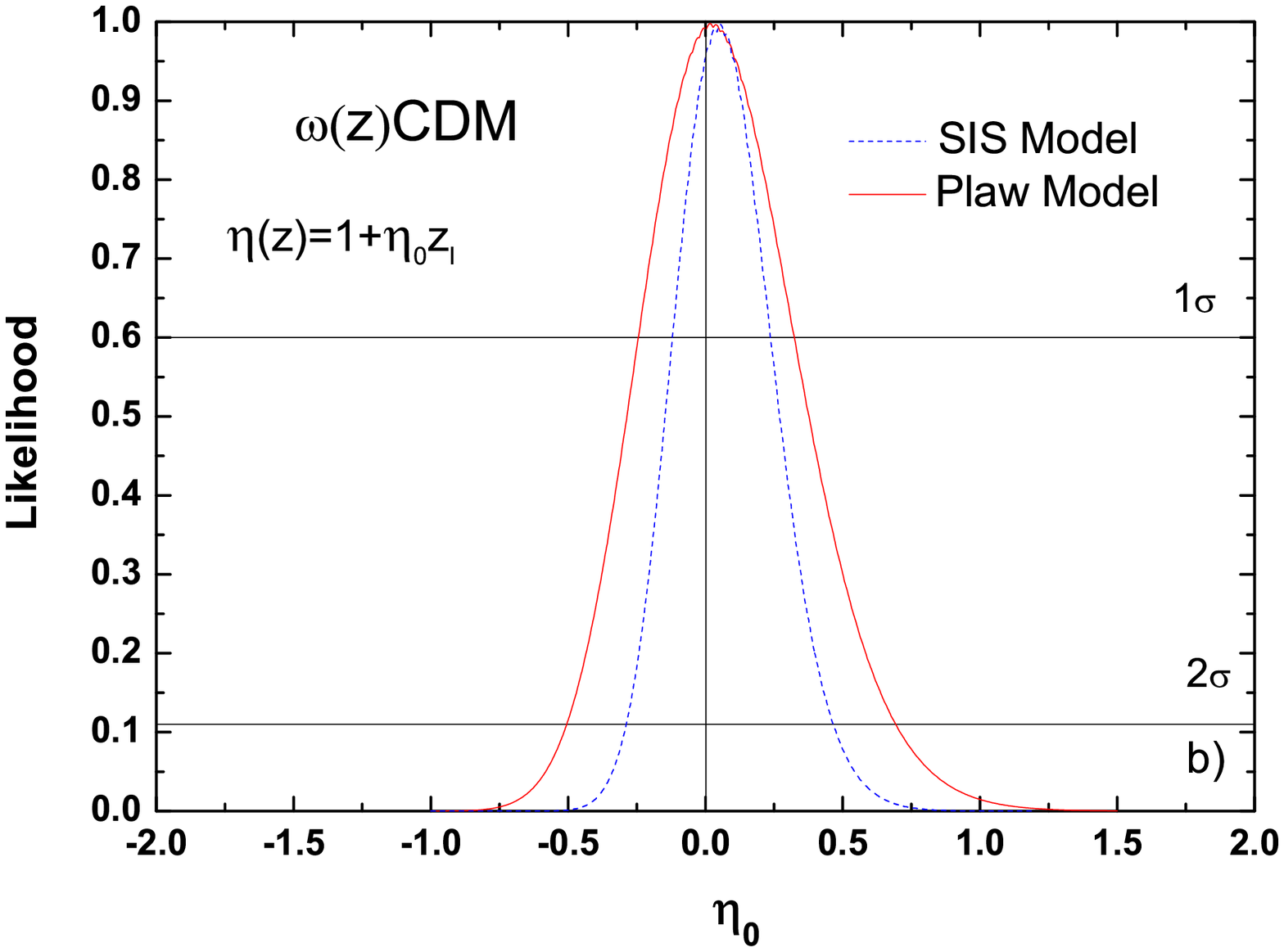}
\caption{The likelihood function for $\eta_0$ parameter in the linear parametrization  assuming the Planck best-fit $\Lambda$CDM model (a) and the WMAP-9 best-fit $\omega (z)$CDM model (b). Note that, regardless of the cosmology assumed, the CDDR validity ($\eta_0 = 0$) is verified at 1$\sigma$.}
\end{figure*}

\begin{figure*}
\centering
\includegraphics[width=0.48\textwidth]{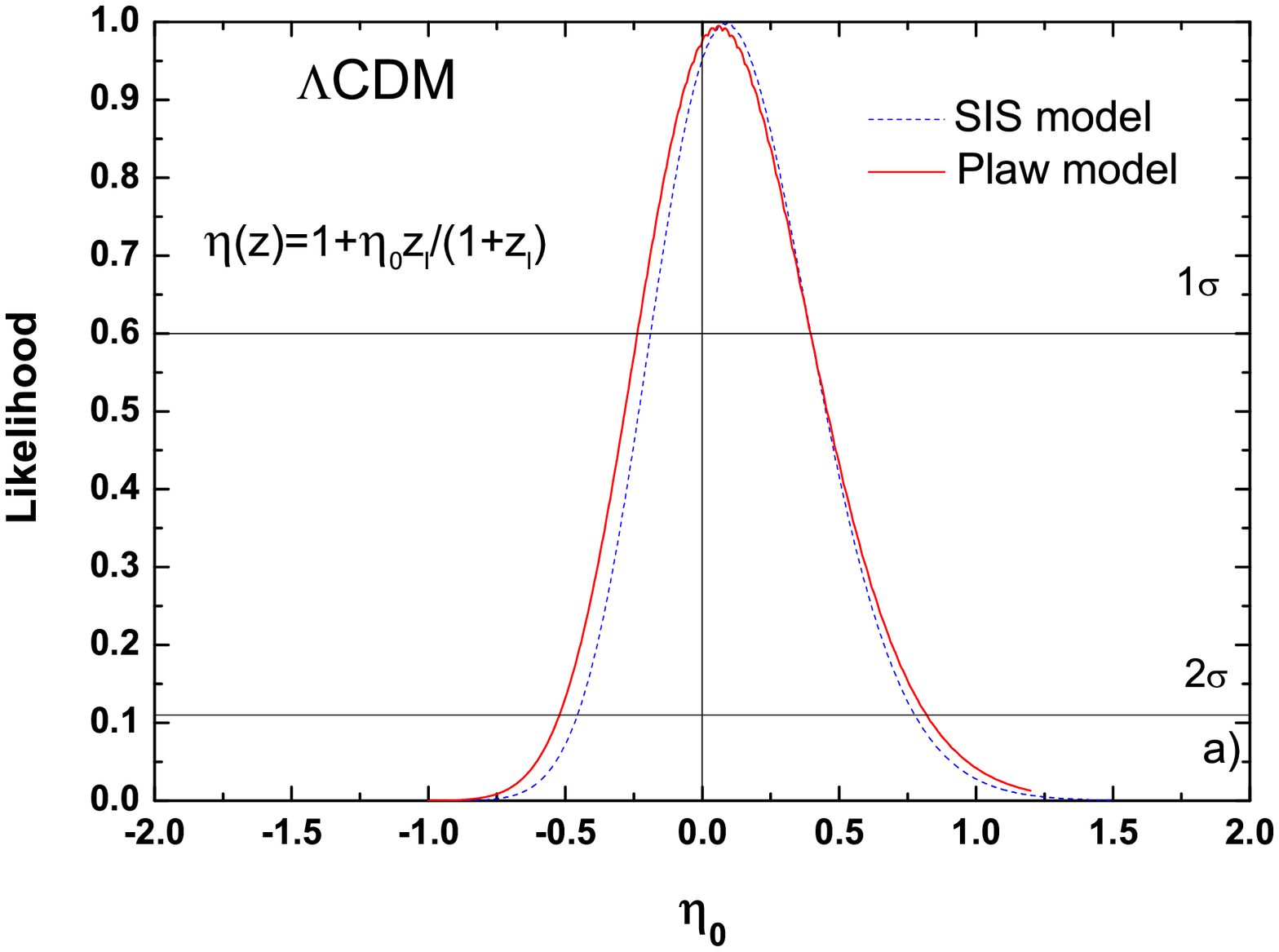}
\hspace{0.3cm}
\includegraphics[width=0.48\textwidth]{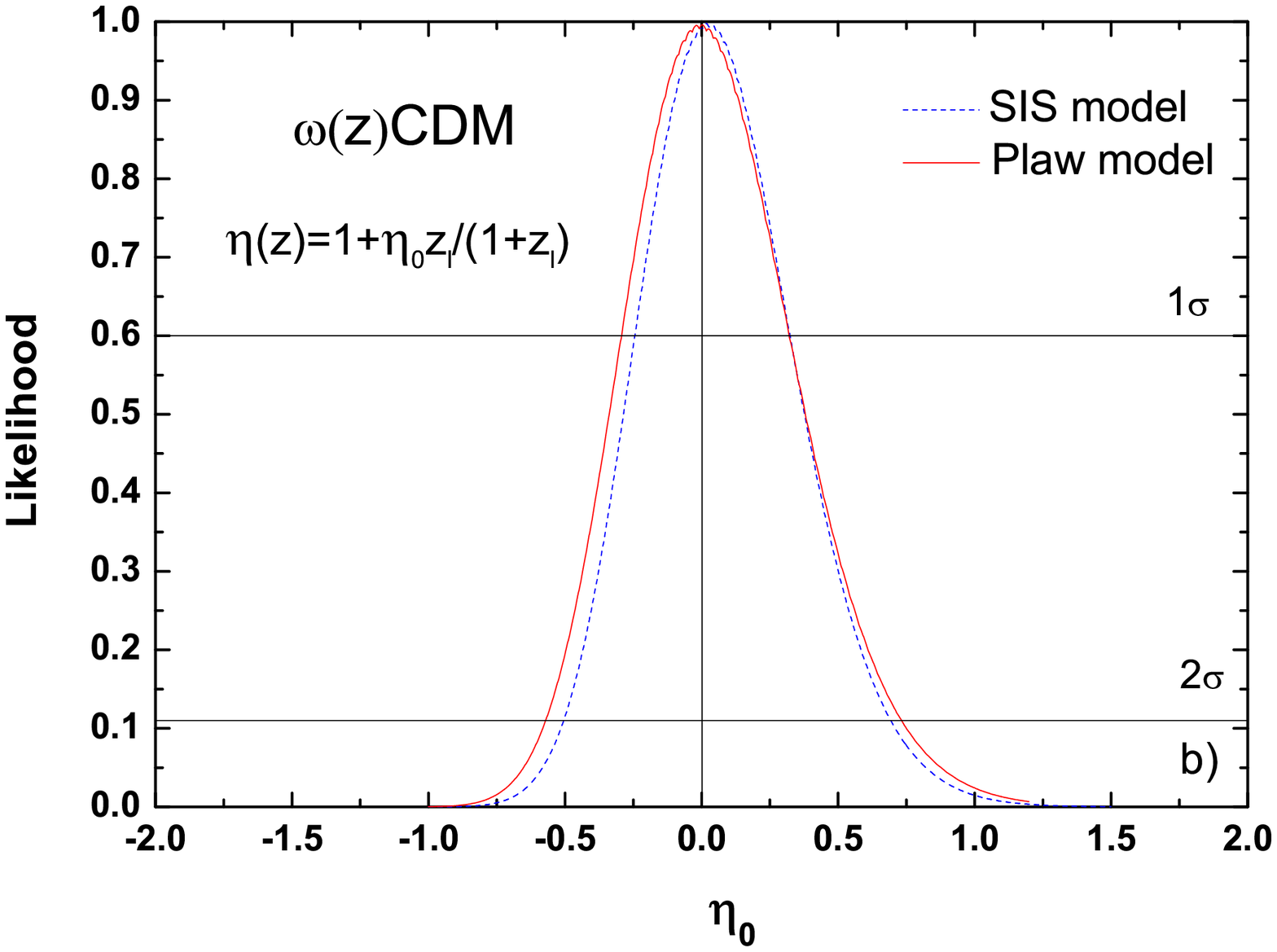}
\caption{The likelihood function for $\eta_0$ parameter in the non-linear parametrization assuming the Planck best-fit $\Lambda$CDM model (a) and the WMAP-9 best-fit $\omega (z)$CDM model (b). Note that, regardless of the cosmology assumed, the CDDR validity ($\eta_0 = 0$) is verified at 1$\sigma$.}
\end{figure*}

\section{Samples}

Recently, the Supernova Cosmology Project reported the discovery of the most distant SNe Ia (Rubinet al. 2013):  the SNe Ia SCP-0401 at $z = 1.713$ and  distance modulus of $45.57 \pm 0.24$ (statistical errors). In our analysis we added this  SNe Ia to Union  2.1 sample (Suzuki et al. 2012), comprising 581 data points. The curves shown in Fig. 1a stand for the  polynomial fit of $D_{\scriptstyle L}$ points from SNe Ia data and the corresponding 1$\sigma$ error. The open circle corresponds to the most distant ($z=1.713$) spectroscopically confirmed SNe Ia. 

We also use 95 data points from  118 SGL systems from Sloan Lens ACS survey (SLACS), BOSS Emission-Line Lens Survey (BELLS), Lenses Structure and Dynamics Survey  (LSD) and  Strong Legacy Survey SL2S surveys (Cao et al. 2015). { We discarded systems with sources
redshifts higher than $z=1.7$}. It is important to stress that Cao et al. (2015) assumed a spherically symmetric mass distribution in lensing galaxies, but relaxed the rigid assumption of SIS model in favor of more
general power-law index $\gamma $ (Plaw), $\rho \propto r^{-\gamma}$, where the distribution becomes a SIS for $\gamma=2$. Under this assumption the Einstein radius is
\begin{equation}
 \label{Einstein} 
\theta_E =   4 \pi
\frac{\sigma_{ap}^2}{c^2} \frac{D_{ls}}{D_s} \left(
\frac{\theta_E}{\theta_{ap}} \right)^{2-\gamma} f(\gamma),
\end{equation}
where $\sigma_{ap}$ is the  stellar velocity dispersion inside the aperture of size $\theta_{ap}$ and
\begin{eqnarray} \label{f factor}
f(\gamma) &=& - \frac{1}{\sqrt{\pi}} \frac{(5-2 \gamma)(1-\gamma)}{3-\gamma} \frac{\Gamma(\gamma - 1)}{\Gamma(\gamma - 3/2)}\nonumber\\
          &\times & \left[ \frac{\Gamma(\gamma/2 - 1/2)}{\Gamma(\gamma / 2)} \right]^2.
\end{eqnarray}
Therefore, 
\begin{equation} 
\label{NewObservable}
 D=D_{A_{ls}}/D_{A_{s}} = \frac{c^2 \theta_E }{4 \pi \sigma_{ap}^2} \left( \frac{\theta_{ap}}{\theta_E} \right)^{2-\gamma} f^{-1}(\gamma).
\end{equation}
The sample used here is compiled and summarized in Table 1 of Cao et al. (2015), in which all relevant information necessary to obtain $D$ from Eqs. (3) and (13) to perform our fit can be found. Also following these authors, we replace $\sigma_{ap}$ by $\sigma_0$ in Eq. (\ref{NewObservable}). This procedure makes $D$  more homogeneous for the sample of lenses located at different redshifts. 

\begin{table*}[ht]
\caption{A summary of the current constraints on the $\eta_0$ and $\epsilon$ parameters from different observables.}
\label{tables1}
\par
\begin{center}
\begin{tabular}{|c||c|c|c|c|c|}
\hline\hline Reference & Data Sample &$1+\eta_0z$ & $1+\eta_0z/(1+z)$& $(1+z)^{2+\epsilon}$ & $\eta(z)=\eta_0$ 
\\ \hline\hline 
Uzan et al. (2004) & ADD\footnote{ADD means angular diameter distance from Galaxy clusters using ESZ/X-ray technique. In all cases, the galaxy clusters were modelled by a elliptical $\beta$ model.} + $\Lambda$CDM & - & - & -& $0.91 \pm 0.04$ \\
Lazkoz et al. (2006)*& SNe Ia + CMB\footnote{Cosmic Background Radiation} + BAO\footnote{Barion acoustic oscillations} &-&-&-& $0.95 \pm 0.025$ \\
De Bernardis et al. (2006) & ADD + $\Lambda$CDM &-&-& -&$0.97 \pm 0.03$ \\
Holanda et al. (2010) & ADD  + $\Lambda$CDM & $-0.056 \pm 0.1$ &$-0.088 \pm 0.14$& - &- \\
Avgoustidis et al. (2010)* & SNe Ia + $\Lambda$CDM + $H(z)$& - & &$ -0.01^{+0.08}_{-0.09}$ & -\\ 
Holanda et al.  (2011)*\footnote{The symbol ``*'' means 2$\sigma$ error bars } & ADD  + SNe Ia  &$-0.28 \pm 0.44$ & $-0.43 \pm 0.66$ & - &- \\ 
Li et al. (2011)*&  ADD  + SNe Ia & $-0.12 \pm 0.35$ & $-0.25 \pm 0.20$ & - &-\\
Avgoustidis et al. (2012)* & SNe Ia + $\Lambda$CDM + $H(z)$&-& - &$ -0.04^{+0.08}_{-0.07}$ &-\\ 
Nair et al. (2012) &BAO + SNe Ia &  $-0.098 \pm 0.084$    & $-0.151 \pm 0.155$ & - &-\\
Gon\c{c}alves et al. (2012)* & Gas mass fractions + SNe Ia  &$-0.03^{+1.03}_{-0.65}$& $-0.08^{+2.28}_{-1.22}$ & - &-\\
Holanda et al. (2012)* & Only Gas mass fractions & $-0.06 \pm 0.16$ & $-0.07 \pm 0.24$ &- & - \\
 Yang et al. (2013) & ADD   + SNe Ia      & $0.16^{+0.56}_{-0.39}$    & - & - &-\\
Holanda et al. (2013) & SNe Ia + $H(z)$ &-&-& $0.017 \pm 0.055$ &- \\
 Jhingan et al. (2014) & Radio galaxies + SNe Ia &  $-0.180 \pm 0.244$ & $-0.415 \pm 0.632$ & - &- \\
Santos-da-Costa et al. (2015)* & ADD  + $H(z)$   & $-0.100^{+0.117}_{-0.126}$ & $-0.157^{+0.179}_{-0.192}$& - &- \\
Santos-da-Costa et al. (2015)* & Gas mass fraction  + $H(z)$ & $0.062^{+0.168}_{-0.146}$ & $-0.166^{+0.337}_{-0.278}$ & - &- \\
Chen et al. (2015)& ADD + SNe Ia + $H(z)$ &$0.07 \pm 0.08$ & $0.15 \pm 0.18$ &- &- \\
Puxon et al. (2015) & SNe Ia + BAO & $-0.027 \pm 0.064$ & $-0.039 \pm 0.099$& -&- \\
Liao et al. (2015) &Str. grav. lens. + SNe Ia & $ -0.004^{+0.322}_{0.210}$ & $ - $ & - &-\\
\bf{This paper}\footnote{Planck results} & SGL (SIS) + SNe Ia + $\Lambda$CDM & $0.05 \pm 0.15$ & $0.09 \pm 0.3$& - &-\\
\bf{This paper}$^{e}$ & SGL (PLaw) + SNe Ia + $\Lambda$CDM & $0.08 \pm 0.22$ & $0.06 \pm 0.33$&  &-\\
\bf{This paper}\footnote{WMAP9 results} & SGL (SIS) + SNe Ia + + $\omega(z)$CDM & $0.01 \pm 0.22$ & $0.017 \pm 0.28$& - &-\\
\bf{This paper}$^{f}$ & SGL (PLaw) + SNe Ia + $\omega(z)$CDM & $0.054 \pm 0.29$ & $0.0035 \pm 0.3$& - &-\\
\hline\hline
\end{tabular}
\end{center}
\end{table*}

\section{Results}

In this work we assume the following general expressions: $ \eta (z_l) = 1 + \eta_{0}z_l$ (linear parametrization) and $ \eta (z_l) = 1 + \eta_{0}z_l/(1+z_l)$ (non-linear parametrization), which  avoids divergences at high-$z$. The constraints to the $\eta_0$ parameter are derived by evaluating the likelihood distribution 
function, ${\cal{L}} \propto e^{-\chi^{2}/2}$, with
\begin{eqnarray}
\chi^{2} & = & \sum_{z_l}\frac{\left[\eta(z_l)-D^{obs}\right]^2}{\sigma^2}                                                                
\end{eqnarray}
where  $D^{obs}$ is the observed $D(1+z_l)^{-2}/D^*$ ratio  and  $\sigma^2$ stands for the statistical errors associated to SNe Ia, 
gravitational lensing 
obtained using standard propagation errors techniques. For the gravitational lensing error one may show that
\begin{equation} \label{uncertainty}
\sigma_D = D \sqrt{4 (\delta \sigma_{0})^2 + (1-\gamma)^2 (\delta \theta_E)^2}\;.
\end{equation}
In order to explore the influence of the model used to describe the lens we perform our analysis by using the Eqs.  (3) and (13) where the following flat priors on $f_e$ and $\gamma$ were used: $0.85 < f_e < 1.15$ and $1.15 < \gamma < 3.5$. {The results of our statistical analysis are shown in Figs. (2) and (3).}

{In Figs. (2a) and (2b) we plot the results  for the linear parametrization. The parameter $\eta_0$ describing possible departures from the CDDR is constrained in the regions: $-0.1 \leq \eta_0 \leq 0.2$ and $-0.14 \leq \eta_0 \leq 0.3$  for $\Lambda$CDM model using  SIS and  Plaw SGL models, respectively.  For $\omega(z)$CDM model we obtain: $-0.21 \leq \eta_0 \leq 0.23$ and $-0.23 \leq \eta_0 \leq 0.35$ using the SIS and Plaw models, respectively.

 In Figs. (3a) and (3b) we plot the results for the non-linear parametrization. The parameter $\eta_0$ is constrained in the regions: $-0.2 \leq \eta_0 \leq 0.4$ and $-0.25 \leq \eta_0 \leq 0.40$  for $\Lambda$CDM model using  SIS and  Plaw SGL models, respectively.  For $\omega(z)$CDM model we obtain: $-0.26 \leq \eta_0 \leq 0.28$ and $-0.3 \leq \eta_0 \leq 0.3$ using the SIS and Plaw models, respectively. As on may see, the validity of the CDDR is fully compatible with the data regardless the $\eta(z_l)$ parametrization or cosmological model adopted in the analysis. Finally, in Table I we show an extensive list of current bounds on a possible departure from the CDDR obtained from different techniques. It is worth observing that the validity of the CDDR  is verified at 2$\sigma$ in all cases.  Although the uncertainties in our analysis are larger than previous works, given the large scatter of distance measurements  in the lensing observations, the advantage of using these systems  are the few astrophysical assumptions and systematic uncertainties.}

\section{Conclusions}

The cosmic distance-duality relation (CDDR), which relates in a simple way  the luminosity and the angular 
diameter distances at a given redshift $z$, plays an essential role in observational cosmology. Its validity has a direct impact on several cosmological probes, from  gravitational lensing studies to the spectrum of the cosmic microwave background and galaxy clusters observations. Any source of attenuation, such as gray intergalactic dust or exotic photon interaction, contributes to violate the
CDDR since its proof is based on the conservation of the average number of photons. On the other hand, deviations from the CDDR can 
also indicate that light does not propagate on null geodesics or that a metric theory for gravitation is inadequate. Therefore, given the importance of the underlying hypotheses which hold the CDDR, tests of the CDDR are an important task for cosmology. 

In this paper we have discussed a new test of the CDDR in terms of the ratio of angular diameter ($D=D_{A_{12}}/D_{A_{2}}$) and luminosity ($D^{*}=D_{L_{12}}/D_{L_{2}}$) distances, where the subscripts 1 and 2 correspond to redshift $z_1$ and $z_2$. 
By allowing departures of the CDDR we have proposed  $D(1+z_1)^{-2}/D^*=\eta(z)$, where $\eta(z)=1+\eta_0{z}/(1+\eta_0(z))$,  and placed constraints on $\eta_0$ by using strong gravitational lensing and  type Ia Supernovae data. More specifically, we used 95 $D$ measurements from various gravitational lens and $D^*$ from 581 SNe Ia (Union 2.1 sample plus the most distant SNe Ia to date, i.e., the SCP-040 at $z = 1.713$).  We have explored the influence of the model used to describe the lens by performing the fits under the assumption of the singular isothermal sphere and singular isothermal sphere with the general power-law index $\gamma$. We have obtained no violation of the CDDR. The results  derived here show an excellent agreement with those originated from very different astrophysical environments  (see Table I) and reinforce
the theoretical pillars of the CDDR.

\acknowledgements

RFLH acknowledges financial support from INCT-A and CNPq (No. 478524/2013-7). JSA is supported by CNPq (No. 310790/2014-0; 400471/2014-0), FAPERJ and INEspa\c{c}o.  VCB is supported by  S\~{a}o Paulo Research Foundation (FAPESP)/CAPES agreement,
under grant number 2014/21098-1.

\label{lastpage}
\end{document}